\documentclass[%
 amsmath,amssymb,
 aps, physrev,
twocolumn,
]{revtex4-2}

\usepackage{graphicx}
\usepackage{dcolumn}
\usepackage{bm}
\usepackage{lipsum}
\usepackage{xcolor}
\usepackage{braket}
\usepackage{siunitx}
\usepackage{mathtools}
\usepackage[colorlinks=true,
            linkcolor=blue,
            citecolor=blue,
            urlcolor=blue]{hyperref}

\newcommand{\emax}{\mathrm{e}_{\mathrm{max}}}
\newcommand{\etmax}{\mathrm{e}_{\mathrm{3max}}}
\newcommand{\nnlogotnf}{{\Delta}{\rm NNLO}_{\rm GO}(394)}
\newcommand{\nnlotnf}{{\Delta}{\rm NNLO}(394)}

\DeclarePairedDelimiter\abs{\lvert}{\rvert}

\begin{document}

\preprint{APS/123-QED}

\title{\textbf{Chiral interactions and superfluidity in the calcium isotopic chain} 
}%

\author{A.~Scalesi}
    \email{alberto.scalesi@chalmers.se}
    \affiliation{Department of Physics, Chalmers University of Technology, SE-412 96 G\"oteborg, Sweden}
\author{A.~Ekstr\"om}
    \affiliation{Department of Physics, Chalmers University of Technology, SE-412 96 G\"oteborg, Sweden}
\author{C.~Forss\'en}
    \affiliation{Department of Physics, Chalmers University of Technology, SE-412 96 G\"oteborg, Sweden}
\author{G.~Hagen}
    \affiliation{Physics Division, Oak Ridge National Laboratory, Oak Ridge, Tennessee 37831, USA}
    \affiliation{Department of Physics and Astronomy, University of Tennessee, Knoxville, Tennessee 37996, USA}

\date{\today}

\begin{abstract}
We perform \emph{ab initio} calculations of three-point mass differences in the odd- and even-mass $^{39-49}$Ca isotopes to probe nuclear superfluidity via empirical neutron pairing gaps. We also quantify the sensitivity of those gaps to the parameters of the interaction at mean-field level.
Recent studies employing accurate chiral nuclear interactions have found these gaps to be too small. 
We show that experimental values can be reproduced at mean-field level by substantially increasing the attraction of the singlet $S$-wave two-nucleon contact interaction, but doing so induces an unphysical bound state of the di-neutron.
The sensitivity of these predictions to the full calibration of the nuclear interaction is then studied by performing Bayesian posterior sampling in a delta-full chiral effective field theory at third chiral order. We find that pairing gaps remain largely unaffected, leaving the explanation of nuclear superfluidity as a future task for improved many-body modeling and refined interactions at higher chiral orders.

\end{abstract}

\maketitle

\paragraph*{Introduction.}
Nuclear superfluidity~\cite{Cooper:1959zz} is an essential ingredient to explain the physics of neutron stars and nuclear matter~\cite{Dean03} as well as nuclear-structure phenomena~\cite{Duguet01a,Duguet01b,Duguet20,Drissi21a,Drissi21b} such as the staggering of nuclear masses, global patterns of first excited states in even- and odd-mass nuclei, and collective aspects of rotating and vibrating nuclei. On the microscopic level, nuclear superfluidity arises from nucleon pairing driven by the strong attractive nuclear interaction. This pairing occurs mainly in the $S$-wave channel and, at higher energies and densities, also in $P$-wave channels. 

A realistic and systematically improvable description of the strong nuclear interaction is provided by chiral effective field theory ($\chi$EFT)~\cite{Hammer20b,Epelbaum09,Machleidt11}, which promises a link between nuclear forces and low-energy quantum chromodynamics. Meanwhile, pairing correlations between nucleons can be partially accounted for at the nuclear mean-field level using Hartree-Fock-Bogoliubov (HFB) theory~\cite{RingSchuck,Schunck19}. This is a variational approximation to the many-body wave function that describes superfluidity by spontaneously breaking the U(1) particle-number symmetry. \emph{Ab initio}~\cite{Ekstrom23} correlation-expansion methods~\cite{Hergert20} enrich such wave functions through the resummation of dynamical correlations~\cite{Signoracci15,Tichai18a}, which can be implemented in a perturbative or nonperturbative fashion, reaching key nuclei across the nuclear chart~\cite{Tichai23}.

\begin{figure}
\includegraphics[width=1\columnwidth]{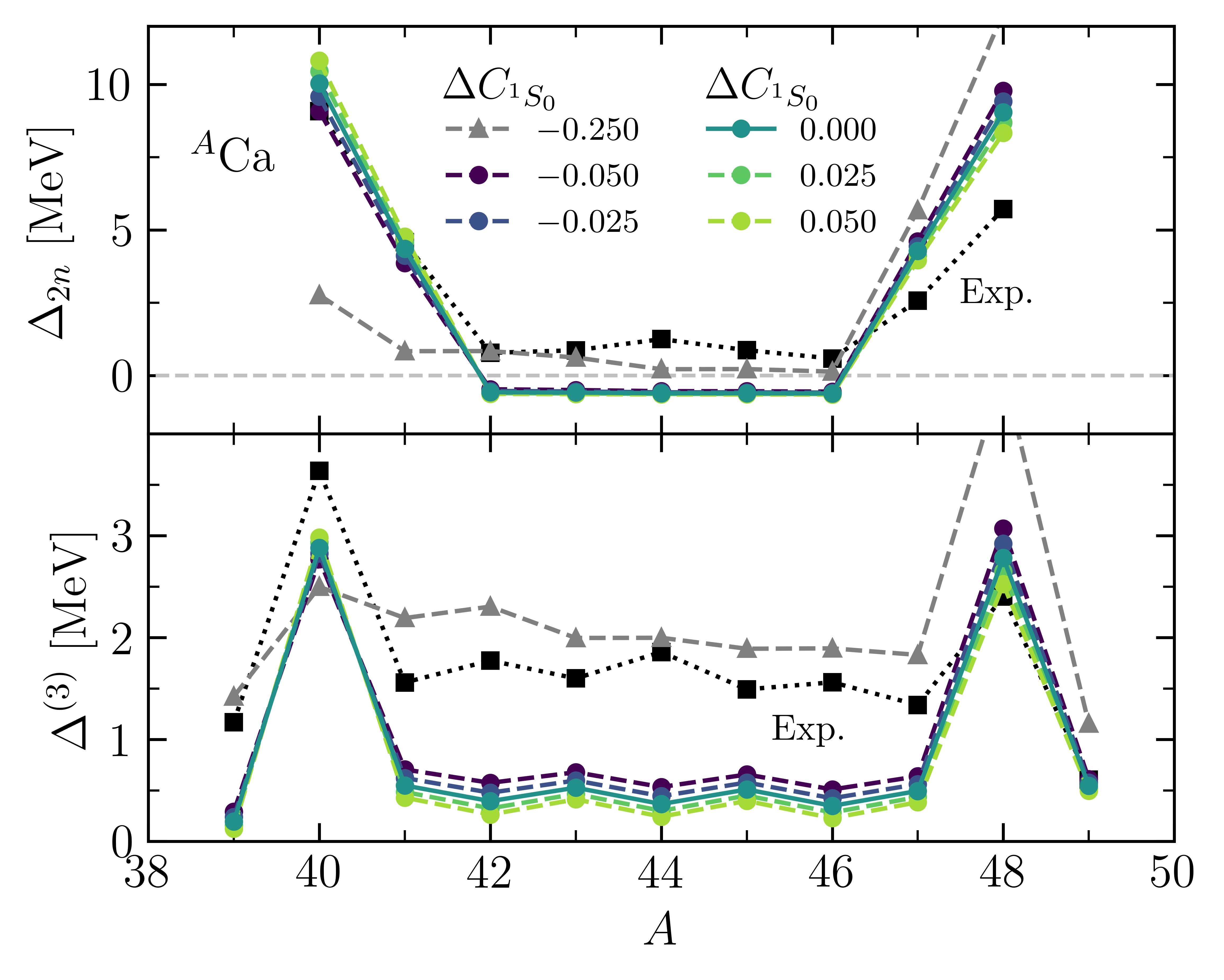}
\caption{\label{fig:C1S0_variations} Impact of the variation of the subleading low-energy constant (LEC) $C_{^1S_0}$ on two-neutron shell gaps $\Delta_{2n}$ (top panel) and three-point mass differences $\Delta^{(3)}$ (bottom panel) computed with the sHFB method. Values in the legend represent the shift $\Delta C_{^1S_0}$ applied to the $\nnlogotnf$~\cite{Jiang20} value of $C_{^1S_0} = 2.505 \times10^4\,\si{\giga\eV^{-4}}$ in the same unit. Experimental data (black squares)~\cite{Wang21} are shown for comparison.}
\end{figure}

Polynomially scaling \emph{ab initio} calculations in the calcium isotopic chain, employing Bogoliubov reference states with the $NN +3N$(lnl)~\cite{Soma20a} and 1.8/2.0 chiral interactions~\cite{Hebeler11}, yield very little of the observed pairing correlations when the many-body expansion is truncated at low orders~\cite{Soma21,scalesi_thesis}. As shown in the present paper, this shortfall is evident when studying the three-point mass difference $\Delta^{(3)}$, defined below and commonly used as a measure of the pairing gap (see solid line in the bottom panel of Fig.~\ref{fig:C1S0_variations}, obtained with the spherical HFB (sHFB) method). The same interactions applied with low-order polynomially-scaling methods reproduce with good accuracy ground-state and two-neutron separation energies in medium-mass nuclei~\cite{Soma20a,Stroberg21}. 

As demonstrated in Ref.~\cite{scalesi_thesis}, VS-IMSRG(2) calculations employing a $^{28}$Si core reproduce the experimental three-point mass differences $\Delta^{(3)}$. Furthermore, angular-momentum--projected coupled-cluster calculations (PCC)~\cite{SunPvt} improve the agreement with experiment by approximately a factor of two. These findings indicate that non-trivial many-body correlations play an important role in a realistic description of these pairing gaps.
While this suggests that the observed reduction of pairing correlations in \emph{ab initio} calculations originates primarily from approximations in the many-body expansion, the role played by chiral interactions remains to be quantified.

In this letter we analyze to which extent the uncertainties associated with the calibration of $\chi$EFT interactions influence the \emph{ab initio} mean-field description of nuclear superfluidity as observed via the magnitude of $\Delta^{(3)}$ in the calcium isotopic chain~\cite{Soma21}. We specifically examine the role of the subleading singlet-$S$ contact LEC $C_{^{1}S_{0}}$, which controls the strength of the dominant $T=1$, $J=0$ pairing channel in nuclei. The value of this LEC has been identified to strongly impact nuclear deformation in $\chi$EFT~\cite{Sun25,becker25}. This is physically well motivated, as pairing correlations and the associated nuclear superfluidity are known to impact the moment of inertia and rotational structure in nuclei~\cite{RingSchuck}.

\paragraph*{Methods and results}
The mean-field calculations presented in this work are primarily performed using the sHFB method, with the inclusion of $3N$ forces~\cite{Signoracci15}. This method should capture the formation of Cooper pairs in finite nuclei by treating the mean field and pairing field self-consistently.
Ground-state energies of odd-even nuclei are computed by generating a fake-odd nucleus with correct average number of particles and adding the smallest quasi-particle energy to its ground-state energy~\cite{Duguet01a}.
A more accurate description of odd-even nuclei would be accomplished by performing a fully-blocked HFB calculation. Such an approach would however be more computationally expensive and is not expected to significantly impact the three-point mass difference, given that even low-order beyond mean-field methods performed on top of an sHFB reference state fail to capture the experimental trend~\cite{Soma21}.
The nuclear interaction is expanded on a spherical harmonic oscillator single-particle basis with oscillator frequency $\hbar\omega = 12$ MeV. The one-body and three-body Hilbert spaces are truncated to $\emax = 12$ and $\etmax = 18$, respectively. We generated nuclear interaction files using the \texttt{NuHamil} code~\cite{Miyagi23}. We employ the $\nnlogotnf$ chiral interaction~\cite{Jiang20} from $\chi$EFT at next-to-next-to-leading order (NNLO), including intermediate $\Delta(1232)$-isobar excitations, as the starting point for our study of the $^{39-49}$Ca isotopes. This interaction provides a realistic description of the ground-state energies up to $^{56}$Ca, but when employed in approximate many-body methods such as in low-rank coupled-cluster, VS-IMSRG, and self-consistent Green's function methods, it fails to explain the parabolic evolution of the charge radii of the open-shell $^{42-46}$Ca isotopes and the increasing charge radii of the neutron-rich calcium isotopes~\cite{GarciaRuiz16} --- a longstanding puzzle~\cite{Caurier01}. A sensitivity study has been performed recently in~\cite{Companys25}, where the leading 3N LECs have been varied to study their impact on charge radii in $^{48-52}$Ca isotopes (see also~\cite{Heinz:2024juw}). This work concluded that neglected many-body correlations are likely the reason for the discrepancy with data.

In this letter we focus on two differential-energy quantities, namely\\
(1) \emph{The two-neutron shell gap}
\begin{equation}
\Delta_{2n}(N) \equiv S_{2n}(N)-S_{2n}(N+2),
\end{equation}
expressed in terms of the two-neutron separation energy $S_{2n}(N) \equiv |E(N)| - |E(N-2)|$, where $E(N)$ denotes the ground-state energy of an isotope with $N$ neutrons and $Z$ protons. This quantity provides information on the location of shell closures. In open shells, it also measures the curvature of the total energy, i.e.~its second derivative with respect to the number of neutrons~\cite{Scalesi24b}.\\
(2) \emph{The three-point mass difference}
\begin{equation}
\Delta^{(3)}(N) \equiv \frac{(-1)^N}{2} [E(N-1)-2E(N)+E(N+1)]
\end{equation}
provides a measure of the empirical pairing gap and the degree of superfluidity in the isotope under study as well as information on the curvature of the energy through the direction of its oscillations~\cite{Duguet01b}.

Both $\Delta_{2n}$ and $\Delta^{(3)}$ computed at mean-field level exhibit quantitatively similar results to low-order beyond mean-field polynomial methods. To validate this we compare in Tab.~\ref{fig:comparisonCCSD} our results from sHFB with deformed axially-symmetric Hartree-Fock (dHF) and deformed coupled-cluster calculations with singles and doubles excitations (dCCSD) starting from a dHF reference state~\cite{Novario20,kortelainen2022,novario2023}, with a rank-reduced three-body interaction~\cite{Frosini21} acting as an effective two-body interaction.
We note that the breaking of rotational symmetry in our coupled-cluster computations should be restored~\cite{Hagen22,Sun25}, and will have an impact on the differential quantities we study in this work. Furthermore, even if in principle allowed by the deformed methods employed, spherical symmetry is not broken for $^{40}$Ca and $^{48}$Ca.

\begin{table*}
\centering
\begin{tabular}{c @{\hspace{25pt}} ccccc @{\hspace{25pt}} ccccc}
 & \multicolumn{5}{c}{$\Delta_{2n}$ [MeV]} 
 & \multicolumn{5}{c}{$\Delta^{(3)}$ [MeV]} \\
\cline{2-6}\cline{7-11}
 & sHFB & dHF & dCCSD & PPD & Exp. 
 & sHFB & dHF & dCCSD & PPD & Exp. \\
\hline
$^{39}$Ca & — & — & — & — & — 
          & 0.19 & $-0.11$ & 0.07 & $0.34^{+0.20}_{-0.30}$ & 1.17 \\
$^{40}$Ca & 10.04 & 10.11 & 10.93 & $10.91^{+0.85}_{-0.83}$ & 9.09
          & 2.88 & 2.64 & 2.94 & $3.21^{+0.30}_{-0.33}$ & 3.64 \\
$^{41}$Ca & 4.36 & 3.71 & 4.47 & $4.88^{+0.70}_{-0.58}$ & 4.58
          & 0.55 & 0.33 & 0.35 & $0.65^{+0.15}_{-0.14}$ & 1.56 \\
$^{42}$Ca & $-0.57$ & $-1.45$ & $-1.11$ & $-0.63^{+0.21}_{-0.30}$ & 0.78
          & 0.39 & $-0.13$ & 0.00 & $0.49^{+0.19}_{-0.19}$ & 1.77 \\
$^{43}$Ca & $-0.59$ & $-0.50$ & $-0.34$ & $-0.64^{+0.25}_{-0.28}$ & 0.87
          & 0.53 & 0.15 & 0.19 & $0.64^{+0.15}_{-0.14}$ & 1.60 \\
$^{44}$Ca & $-0.61$ & 0.08 & 0.11 & $-0.65^{+0.25}_{-0.30}$ & 1.25
          & 0.37 & 0.17 & 0.22 & $0.47^{+0.20}_{-0.20}$ & 1.86 \\
$^{45}$Ca & $-0.61$ & 0.16 & 0.24 & $-0.64^{+0.27}_{-0.31}$ & 0.87
          & 0.51 & 0.15 & 0.19 & $0.62^{+0.15}_{-0.14}$ & 1.49 \\
$^{46}$Ca & $-0.61$ & $-0.15$ & $-0.03$ & $-0.64^{+0.26}_{-0.35}$ & 0.59
          & 0.35 & 0.22 & 0.28 & $0.46^{+0.22}_{-0.21}$ & 1.56 \\
$^{47}$Ca & 4.28 & 4.07 & 4.37 & $2.96^{+0.68}_{-0.67}$ & 2.57
          & 0.49 & 0.36 & 0.39 & $0.61^{+0.16}_{-0.15}$ & 1.34 \\
$^{48}$Ca & 9.05 & 8.88 & 9.54 & $6.60^{+1.47}_{-1.58}$ & 5.72
          & 2.78 & 2.54 & 2.68 & $2.21^{+0.32}_{-0.37}$ & 2.40 \\
$^{49}$Ca & — & — & — & — & —
          & 0.55 & 0.27 & 0.21 & $0.57^{+0.07}_{-0.08}$ & 0.61 \\
\end{tabular}
\caption{Comparison between dHF, sHFB, dCCSD theoretical calculations and experimental data for $\Delta_{2n}$ and $\Delta^{(3)}$.
The PPD columns list the mean and 68\% degree of belief intervals of the posterior predictive distributions shown in Fig.~\ref{fig:violin_plot}. The credible region for $\Delta_{2n}$ of $^{40,48}$Ca provides a model check as these two observables entered our likelihood calibration. Experimental ground-state energies are taken from~\cite{Wang21}.}
\label{fig:comparisonCCSD}
\end{table*}
The observed differences between the mean-field and dCCSD results are negligible when compared to the discrepancy with respect to experiment, which confirms the necessity of including more advanced types of collective correlations in the beyond mean-field expansion for a proper description of the gaps.
In order to probe the sensitivity of such quantities on the parameters of the chiral interactions, we restrict our analysis at the mean-field level using sHFB.
This allows for a substantial reduction in computational cost, enabling the large number of calculations presented in this work to be performed. A study of the impact of variations of LECs combined with highly non-perturbative methods (e.g.~VS-IMSRG or PCC) remains an interesting question for future investigation.

Chiral interactions are parametrized by a set of low-energy constants (LECs) that encapsulate unresolved physics. The third-order $\Delta$-full interactions we employ include 17 LECs in total, with 4 LECs $c_1,c_2,c_3,c_4$ in the subleading pion-nucleon ($\pi$N) sector, 2 LECs $c_D,c_E$ that govern the contact parts of the leading 3N force, and the remaining 11 LECs belong to the $NN$ contact potential. The values of the LECs are typically calibrated using experimental data from $NN$ and $\pi$N scattering and ground-state energies and charge radii of selected nuclei. 
We seek to analyze the impact of LEC uncertainty on shell- and pairing gaps of calcium isotopes.

Considering that nuclear Cooper pairs, which generate nuclear superfluidity, typically form in the $^1S_0$ channel at low densities, we will begin by studying the isolated impact of the subleading $NN$ contact LEC $C_{^1S_0}$. This directly determines the interaction strength in the $^1S_0$ channel and consequently also governs $NN$ pairing at low energy. The relation to the pairing gap can be illustrated analytically by considering the well-known weak-coupling limit of the BCS gap equation~\cite{Bardeen57}:
\begin{equation}
\Delta \approx \exp{\biggl( -\dfrac{1}{\abs{V_{\rm eff}}\rho(E_F)} \biggr)},
\end{equation}
where $V_{\rm eff}$ is the effective pairing-interaction between two particles with opposite momenta near the Fermi surface, and $\rho(E_F)$ is the density of states at the Fermi surface. In this approximation, the interaction is taken as constant and attractive near the Fermi surface, which is justified when the pairing strength is small compared to the typical single-particle level spacing around the Fermi surface. For the present argument, the proportionality $V_{\rm eff} \sim C_{^1S_0}$ is assumed.
A smaller (i.e., less positive) value of $C_{^1S_0}$ increases $NN$ attraction and therefore also produces a larger pairing gap---thus enhancing superfluidity. In practice, the interaction is regularized and also includes pion-mediated long-range components; however, $C_{^1S_0}$ governs only the short-range physics that is crucial for pairing. At low nuclear densities, where long-range contributions are suppressed, the dominance of $C_{^1S_0}$ becomes even more pronounced.

In light of this simple picture we investigate the impact of the $C_{^1S_0}$ LEC on $\Delta_{2n}$ and $\Delta^{(3)}$ at mean-field level by varying the nominal $\nnlogotnf$ value, $C_{^1S_0} = +2.505\times 10^{4}$ GeV$^{-4}$, while keeping the other LEC values fixed. The results are shown in Fig.~\ref{fig:C1S0_variations}. We find that percent-level variations $\Delta C_{^1S_0}$ in the range $[-0.050, 0.050] \times 10^4$ GeV$^{-4}$ have little impact. However a 10\% reduction in the magnitude of this LEC---making the interaction more attractive---shifts the sHFB results for $\Delta^{(3)}$ such that we can reproduce experimental data, see triangle symbols in Fig.~\ref{fig:C1S0_variations}. We also note that the oscillation of $\Delta^{(3)}$ across the calcium isotopes follows the experimental data up to $^{43}$Ca. At the same time, the results for $\Delta_{2n}$ indicate a too small shell closure near $^{40}$Ca, and for $^{48}$Ca it becomes too large compared to experiment. Also, the overly strong $NN$ attraction produces an unphysical bound di-neutron with a binding energy of $\SI{66}{\kilo\eV}$, obtained from exact diagonalization. The observed enhanced pairing arising from a stronger $S$-wave interaction is consistent with the sensitivity study of Ref.~\cite{Sun25}, which reported a positive correlation between $C_{^1S_0}$ and the ratio $R_{42}=E(4^{+}_1)/E(2^{+}_1)$, an axial-rotor indicator of deformation, in neutron-rich neon and magnesium isotopes. A larger (smaller) $C_{^1S_0}$ leads to a larger (smaller) $R_{42}$, reflecting weaker (stronger) pairing. Indeed, increased pairing yields irrotational flow, which reduces the moment of inertia relative to the rigid-rotor limit.

Given the observed sensitivity to variations in $C_{^1S_0}$ alone, we now turn to exploring the simultaneous variation of all parameters in $\Delta$-full $\chi$EFT at NNLO.
With the advent of fast and accurate emulators~\cite{Duguet:2023wuh} plus relevant error models, Bayesian methods can now be used for parameter inference in $\chi$EFT~\cite{Svensson:2021lzs,Svensson:2023twt,Svensson23,Wesolowski:2021cni,Hu22,Jiang24b}, resulting in posterior probability densities for LEC values. 
\begin{figure*}[ht]
\includegraphics[width=1.8\columnwidth]{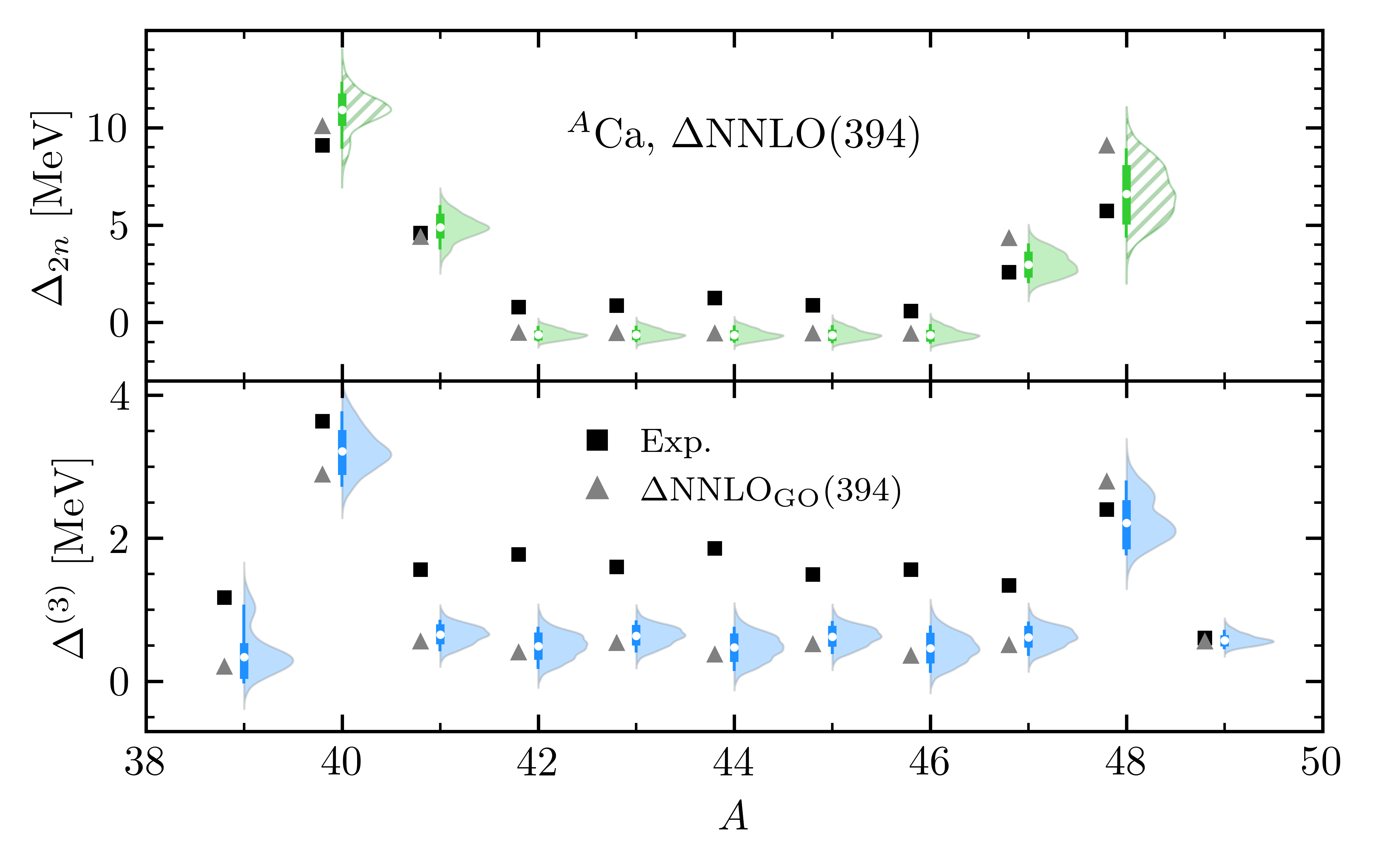}
\caption{\label{fig:violin_plot} Posterior predictive distributions $p( \Delta_{2n}, \Delta^{(3)} \vert \mathcal{D}_\mathrm{cal}, I)$ for calcium isotopes computed with $\nnlotnf$ interactions. White dots indicate the center of the distribution, thick and thin vertical lines enclose $68\%$ and $90\%$ of the probability mass, respectively. Results for the $\nnlogotnf$ interaction and experimental data are shown with triangle and square symbols for comparison. The hatched distributions for $\Delta_{2n}$ of $^{40,48}$Ca provide a model check as these two observables entered our likelihood calibration.}
\end{figure*}
In this work, we use interaction samples from an earlier Bayesian study of nuclear-matter saturation and symmetry energy within $\Delta$-full $\chi$EFT~\cite{Jiang24,Jiang24b}. Our finite set of prior samples has been filtered out in two steps: first, iterative history matching~\cite{Vernon10,Vernon18,Hu22,Kondo23,Jiang24b} was used in \cite{Jiang24,Jiang24b} to find 8192 samples that reproduce (within a non-implausibility window) selected $S$- and $P$-wave $NN$ scattering phase shifts, deuteron properties, the binding energies of $^3$H and $^4$He, and the charge radius of $^4$He. Second, following the same study, a likelihood encompassing the quadrupole moment of the deuteron plus binding energies and radii of $^{2,3}$H, $^4$He, and $^{16}$O was set up and the 164 samples with the largest likelihood weights were retained. Under the approximation that all other samples have negligible posterior probability, this set is then used within the framework of sampling/importance resampling~\cite{Smith:1992aa,Jiang22} to estimate a posterior predictive distribution for $\Delta_{2n}$ and $\Delta^{(3)}$ in the calcium chain of isotopes. 

We first perform sHFB calculations of $\Delta_{2n}$ and $\Delta^{(3)}$ for all 164 interactions in this set. This ensemble of calculations spans a substantial range of values for the $C_{^1S_0}$ LEC, $[2.18, 2.80]\times 10^4~\text{GeV}^{-4}$. To quantify a posterior predictive distribution using the ensemble of sHFB calculations, we proceed as follows. First, we assign likelihood weights $w_i = p(\mathcal{D}_\mathrm{cal} \vert \alpha_i, I)$ using experimental $\Delta_{2n}$ values for $^{40}$Ca and $^{48}$Ca as calibration data $\mathcal{D}_\mathrm{cal}$, with $\alpha_i$ denoting a vector of LECs values from the ensemble. We adopt a normal likelihood with independent errors for the model predictions of the two calibration data. These errors account for uncertainties arising from the sHFB method and EFT truncation, which we estimate to be $\SI{1}{MeV}$ and $\SI{0.5}{MeV}$, respectively. Both error assignments correspond to one standard deviation. The magnitude of the method error is motivated by the study presented in Ref.~\cite{scalesi_thesis}, while the EFT truncation error assignment is consistent with earlier estimates for energy differences~\cite{Kondo23,Sun25}. Model-space truncation errors are found to be negligible, given the large space employed and the use of differential quantities for calibration. 

Finally, a finite-sampling approximation of the posterior predictive distribution $p( \Delta_{2n}, \Delta^{(3)} \vert \mathcal{D}_\mathrm{cal}, I)$ for all calcium isotopes is obtained via importance resampling using the weights $q_i = w_i / \sum_{j=1}^{n} w_j$. The effective number of samples is $n_{\rm eff} = \sum_{i=1}^{n} q_i / q_{\rm max} \approx 10$ and the resulting distribution is smoothed using a Gaussian kernel density estimator. The posterior predictions are shown in Fig.~\ref{fig:violin_plot}. Note that these posteriors reflect parametric uncertainty. EFT truncation and many-body uncertainties affect the analysis only through the calibration likelihood. We find that both $\Delta_{2n}$ and $\Delta^{(3)}$ exhibit small variations relative to the $\nnlogotnf$ reference, indicating that shell closures are stable with respect to the calibration of the nuclear interaction and that nuclear superfluidity is only weakly affected by the EFT parametric uncertainty.

\paragraph*{Summary and outlook.}
We analyzed the impact of parametric uncertainties in a $\Delta$-full chiral interaction on the two-neutron shell gap $\Delta_{2n}$ and the three-point mass-difference $\Delta^{(3)}$ in calcium isotopes calculated at mean-field level. First, we found that the $NN$ contact $C_{^1S_0}$ can have a large influence, in particular when varied while keeping all other LECs fixed. The impact of this LEC is understood, from a simple physical argument, as a dominant contributor to nuclear superfluidity. A 10\% negative shift of $C_{^1S_0}$ from the $\nnlogotnf$ value brought the calcium pairing gap closer to experimental data. However, this was achieved at the cost of producing a bound di-neutron state and significantly altering shell closures. We then examined the impact of simultaneously varying all 17 LECs of the chiral $NN+3N$ interaction, starting from a large set of non-implausible interactions and using importance resampling to obtain the posterior predictive distributions for the relevant observables. From this we concluded that the parametric uncertainties in the third-order chiral interaction have minimal impact on nuclear superfluidity, and cannot remedy the lack of pairing observed in \emph{ab initio} predictions of nuclei. 

Future work should analyze the influence of many-body approximations and symmetry breaking/restoration on nuclear superfluidity. Our estimate of method errors was based on comparisons with available beyond-mean-field methods.
Missing collective correlations not included in these approximations may contribute significantly and explain the discrepancy with the data~\cite{Soma21,Duguet23}.
Past studies~\cite{Caurier01} have shown that an accurate description of the ${\rm BE}(2)$ electromagnetic transition is needed to correctly reproduce the parabolic trend displayed by charge radii in the calcium isotopic chain. It will be interesting to determine whether this quantity is also correlated with nuclear superfluidity and the extent to which it could serve as a possible indicator of this property. The computation of the ${\rm BE}(2)$ will therefore be a pivotal step towards improving agreement with experiment. Moreover, with \emph{ab initio} methods increasingly applied to heavy nuclei~\cite{Stroberg21,Miyagi22,Hu22,Tichai23,Tichai23}, it is of particular interest to determine how such fundamental nuclear quantities can be captured while retaining low-order polynomial scaling in the many-body approach.

\paragraph*{Acknowledgments.}
The authors thank T.~Duguet and V.~Som\`a for useful comments on the manuscript.
This work is inspired by discussions at the ESNT workshop `Ab initio many-body calculations: where has the nuclear pairing gone?' (Saclay, 2025).
A.~S.~thanks H.~Hergert for providing benchmarks for the implementation of three-body forces in the HFB code. This work was supported by the Swedish Research Council (Grants No.~2020-05127, No.~2021-04507, and No.~2024-04681). This work was also supported by the U.S.~Department of Energy (DOE), Office of Science, under SciDAC-5 (NUCLEI collaboration). Computer time was provided by the Innovative and Novel Computational Impact on Theory and Experiment (INCITE) program. This research used resources from the Oak Ridge Leadership Computing Facility located at ORNL, which is supported by the Office of Science of the Department of Energy under Contract No.~DE-AC05-00OR22725.


\bibliography{biblio}

\end{document}